\documentclass[twocolumn,floatfix,superscriptaddress,amsmath,showpacs,showkeys,aps,prb]{revtex4}

\usepackage{t1enc}
\usepackage[final]{graphicx}
\usepackage{graphicx}
\usepackage{epsfig}
\usepackage{bm}
\usepackage[normalem]{ulem}
\usepackage[usenames]{color}

\begin{document}

\title{Inverse transition in the  two dimensional dipolar frustrated ferromagnet}

\author{Sergio A. Cannas}
\email{cannas@famaf.unc.edu.ar}
\affiliation{Facultad de
Matem\'atica, Astronom\'{\i}a y F\'{\i}sica, Universidad Nacional
de C\'ordoba, Instituto de F\'{\i}sica Enrique Gaviola (IFEG-CONICET)\\ Ciudad Universitaria, 5000 C\'ordoba, Argentina}
\author{Marianella Carubelli}
\email{marianela.carubelli@gmail.com}
\affiliation{Facultad de
Matem\'atica, Astronom\'{\i}a y F\'{\i}sica, Universidad Nacional
de C\'ordoba, Instituto de F\'{\i}sica Enrique Gaviola (IFEG-CONICET)\\ Ciudad Universitaria, 5000 C\'ordoba, Argentina}
\author{Orlando V. Billoni}
\email{billoni@famaf.unc.edu.ar}
\affiliation{Facultad de
Matem\'atica, Astronom\'{\i}a y F\'{\i}sica, Universidad Nacional
de C\'ordoba, Instituto de F\'{\i}sica Enrique Gaviola (IFEG-CONICET)\\ Ciudad Universitaria, 5000 C\'ordoba, Argentina}
\author{Daniel A. Stariolo}
\email{daniel.stariolo@ufrgs.br}
\affiliation{Departamento de F\'{\i}sica,
Universidade Federal do Rio Grande do Sul and
National Institute of Science and Technology for Complex Systems\\
CP 15051, 91501-970 Porto Alegre, RS, Brazil}
\altaffiliation{Research Associate of the Abdus Salam International Centre for
Theoretical Physics, Trieste, Italy}

\date{\today}

\begin{abstract}
We show that the mean field phase diagram of the dipolar frustrated ferromagnet in an
external field presents an inverse transition in the field-temperature plane. The
presence of this type of transition has recently been observed experimentally in
ultrathin films of Fe/Cu(001). We study a coarse-grained model Hamiltonian in two
dimensions. The model supports stripe and bubble equilibrium phases, as well as the
paramagnetic phase. At variance with common expectations, already in a single mode
approximation, the model shows a sequence of paramagnetic-bubbles-stripes-paramagnetic
phase transitions upon lowering the temperature at fixed external field. Going beyond
the single mode approximation leads to the shrinking of the bubbles phase, which is
restricted to a small region near the zero field critical temperature. Monte Carlo
simulations results with a Heisenberg model  are consistent with
the mean field results.
\end{abstract}


%75.70.Ak 	Magnetic properties of monolayers and thin films
%75.70.Cn 	Magnetic properties of interfaces (multilayers, superlattices, heterostructures)
%75.60.Jk 	Magnetization reversal mechanisms
%75.70.-i 	Magnetic properties of thin films, surfaces, and interfaces
% 75.30.Kz	Magnetic phase boundaries (including classical and quantum magnetic transitions, metamagnetism, etc.)
%75.70.Kw	Domain structure (including magnetic bubbles and vortices)
\maketitle

\section{Introduction}
\label{Intro}
It is well known that dipolar forces in ferromagnetic systems favor the appearance of domain
structures~\cite{HuSc1998}. In particular, in thin ferromagnetic films with uniaxial out  of plane
anisotropy two kind of patterns are usually observed, stripes and bubbles~\cite{PoVaPe2003,
ChWuWoWuScDoOwQi2007,SaLiPo2010}. Stripes are the stable structures at low temperatures in absence of
external fields~\cite{YaGy1988,AbKaPoSa1995}. When an external magnetic field is present,
a bubble phase may appear in the field-temperature phase diagram~\cite{GaDo1982}.
Regarding the $h-T$ mean field phase diagram,  Garel and Doniach~\cite{GaDo1982} using a single
mode approximation obtained that, as the external field is increased,  a sequence of stripes-bubbles-paramagnet phases have
successively the lower free energy below the critical temperature.  Furthermore,
the transition lines stripes-bubbles and bubbles-paramagnet decay monotonically with growing
temperature, until both meet at a critical point $T_c$ at $h=0$. This behavior was partially
confirmed by simulations of a dipolar lattice gas~\cite{HuSi1992,StSi2002}
and more recently in the dipolar Ising model~\cite{DiMu2010}. Interestingly, this
seemingly established picture of the mean field phase diagram was recently put in question, both from
experimental results on ultrathin films of Fe/Cu(001)~\cite{SaLiPo2010,SaRaViPe2010} and also
from a theoretical point of view~\cite{PoGoSaBiPeVi2010}. The experiments on Fe/Cu(001) show
convincingly an inverse transition sequence paramagnet-modulated-paramagnet when ciclying
in temperature at fixed external applied field. In a recent work, Portmann et. al.~\cite{PoGoSaBiPeVi2010}
addressed the question about the origin of this inverse symmetry breaking transition in
the context of a scaling hypothesis proposed by the authors. The critical field line $h_c(T)$,
which separates the modulated from the paramagnetic phases is a consequence of the balance
between the dipolar energy, which favors the presence of domains, and the external field energy,
which favors a uniform state. Within the proposed scaling form for the energies, this immediately
implies that $h_c \propto M(T)/\lambda(T)$, where $M(T)$ and $\lambda(T)$ are the saturation magnetization inside a
domain  and the modulation length at zero field respectively. As the temperature grows, the magnetization is
first nearly constant while the modulation length strongly diminishes, but on approaching the
critical temperature $M$ decreases faster than $\lambda$. Then the critical field line $h_c(T)$
first grows, passes through a maximum, and finally decays to zero at $T_c$. This means that the
phase diagram has a dome-like shape, instead of the monotonic behavior predicted by Garel and Doniach work\cite{GaDo1982}. This behavior was anticipated by Abanov {\it et al.}~\cite{AbKaPoSa1995} and, in a different context,
 by Andelmann {\it et al.}~\cite{AnBrJo1987}, but no further analysis of their origin or implications was done.

Motivated by these new results, we have reconsidered the mean field phase diagram of a coarse-grained
Hamiltonian for the dipolar frustrated ferromagnet. We have gone beyond the usual single mode
approximation, by considering the effects of higher harmonics in the modulation profiles. This is
expected to be crucial in the context of the recently proposed scaling hypothesis.
The single mode approximation is valid very near the critical point. In
particular, in this approximation, the modulation length is independent of temperature,  at variance with the
experimental observed behavior. Furthermore, in this condition the scaling hypothesis is not expected to be
applicable and a dome-shape of the phase diagram is not to be expected.  Actually,
{\em even in the single mode approximation}, the mean field critical field line passes through a maximum and an inverse transition
is obtained,  as was shown by Andelman {\it et al.}~\cite{AnBrJo1987} (even when they studied a different system, both problems are equivalent,
as shown in the supplementary material), and verified by ourselves.
Going beyond the single mode approximation we determined the variation of the
modulation length with temperature, and show that a few modes are enough in order to get an
asymptotic behavior. Our results are in agreement with previous ones~\cite{PoVaPe2006,ViSaPoPePo2008}.
The most notable effect of including  higher harmonics is the shrinking of the bubbles phase in the
$h-T$ plane. The triple point, already present in the single mode approximation, shifts to
higher temperature, and the whole region where the bubble phase is the thermodynamically stable
one is considerably reduced near the critical temperature. The overall results seem to indicate
a lose of stability of the bubbles phase at low temperatures. These results are compared with
Monte Carlo simulations of a Heisenberg model with exchange and dipolar interactions. A tentative
phase diagram is presented, which is in qualitative agreement with some of the mean field predictions.
In particular, it was not possible to find bubbles at low temperatures for the parameters values studied.
At higher temperatures bubbles seem to be the stable phase for not too low fields.

The plan of the paper is as follows: in Section \ref{mf} we introduce the mean field model. In \ref{mf}A
we compute the profiles of the modulated solutions and the variation of the stripe width with
temperature by considering higher harmonics in the variational solution at zero external field.
In \ref{mf}B we present the results for the $h-T$ phase diagram.
In Section \ref{mc} we show results of Monte Carlo simulations of a Heisenberg model and
compare them with the mean field calculations. In Section \ref{conc} we present our conclusions.

\section{Mean Field Phase Diagram}
\label{mf}

The mean field model is defined by the Landau-Ginzburg free energy:

\begin{widetext}
\begin{equation}
F[\phi] = \frac{1}{2} \int d^2 {\bf x } \left\{ \left( \nabla
\phi({\bf x}) \right)^2 + r_0 \phi^2({\bf x}) + \frac{u}{2}
\phi^4({\bf x}) \right\} + \frac{1}{2\delta} \int d^2 {\bf x} \int
d^2{\bf x}' \phi({\bf x})\phi({\bf x'})J'(\left|{\bf x}-{\bf
x}'\right|) - \int d^2 {\bf x
}\; h({\bf x}) \phi({\bf x}) \label{Hreal}
\end{equation}
\end{widetext}

\noindent where the scalar field $\phi({\bf x})$ represents, {\it e.g.}, the out-of-plane magnetization density
in a magnetic thin film with perpendicular anisotropy. The terms between brackets model  ferromagnetic
behavior in the continuum limit, where $r_0 \propto T-T_F$, $T_F$ being the Curie temperature.
$J'(\left|{\bf x}-{\bf x}'\right|)$  is a (translational invariant) competing interaction, which introduces
frustration and is responsible for modulated patterns at low temperatures. It may represent, {\it e.g.}, the
dipolar interaction in an ultrathin film $J'(r) \propto 1/r^3$. $\delta$ is the ratio between the
ordering ({\it e.g.}, exchange) and the competing ({\it e.g.}, dipolar) interaction intensities and $h({\bf x})$ is an
 external field.  In the Fourier representation the Landau-Ginzburg free energy reads:

\begin{widetext}
\begin{equation}
F[\phi] = \frac{1}{2} \sum_{\bf k} A(k) \phi({\bf k}) \phi({\bf
-k}) + \frac{u}{4\, L^2}  \sum_{{\bf k}_1} \sum_{{\bf k}_2}
\sum_{{\bf k}_3} \phi({\bf k}_1) \phi({\bf k}_2)  \phi({\bf k}_3)
\phi(-{\bf k}_1-{\bf k}_2-{\bf k}_3)- \sum_{\bf k} h({\bf k})\,\phi({-\bf k})\label{Hk}
\end{equation}
\end{widetext}

\noindent where

\begin{equation}
 \phi({\bf x}) =\frac{1}{L} \sum_{{\bf k}}  \phi({\bf k}) e^{i {\bf k}.{\bf x}}
  \label{fourier1}
\end{equation}

 \begin{equation}
A(k) = r_0 + k^2 + \frac{J(k)}{\delta} \label{Ak}
\end{equation}

\noindent and

 \begin{equation}\label{uniform}
J(k) = \int d^2 {\bf r}\ J'(r)\ e^{-i {\bf k}.{\bf r}}
\end{equation}

 The equilibrium configuration is  given by the set of amplitudes $\{\phi({\bf k}) \}$ that minimize the free energy (\ref{Hk}). We will assume that $J(k)$ is such that $A(k)$ has a single minimum at $k=k_m$. At high enough temperatures and in the absence of an external field, the uniform configuration (paramagnetic)  $\phi({\bf k})=0$ is the absolute minimum. The critical temperature at which the uniform solution becomes unstable is given by the condition $A(k_m)=0$. Hence, without loss of generality we can assume the general form

\begin{equation}\label{Ak2}
    A(k) = -b\tau+ c\;(k-k_m)^2
\end{equation}

\noindent  where $\tau\equiv(T_c-T)/T_c$ and $b,c > 0$. This form should be valid close enough of $T_c$. We will consider a uniform external field $h$, so that $h({\bf k}) = L\, h\, \delta_{{\bf k},0}$. For large enough values of $h$, a uniform solution $\phi({\bf k}) = 2\, L\, m_0\, \delta_{{\bf k},0}$, with

\begin{equation}
    2A(0)\, m_0 + 8\, u\, m_0^3=h\label{meanfield-uniform},
\end{equation}

\noindent is expected to be the absolute minimum. For low values of $h$ and
 below the critical temperature $\tau>0$  one expects a modulated solution with a characteristic wave vector $k = k_0$ to become the absolute minimum, where $k_0\to k_m$ when $\tau\to 0$.  We will consider two different types of variational solutions: a striped configuration

\begin{equation}
    \phi_s({\bf x}) = 2 \sum_{n=0}^\infty m_n \cos (n\, {\bf k}_0.{\bf x}),
\end{equation}

\noindent and a bubble type configuration

 \begin{equation}
    \phi_b({\bf x}) = 2 \sum_{n=0}^\infty m_{n} \sum_{i=1}^3 \cos (n\, {\bf q}_i.{\bf x}),
\end{equation}

\noindent where ${\bf q}_i$ are three vectors satisfying $|{\bf q}_i|=k_0$ and $\sum_{i=1}^3 {\bf q}_i =0$.
The Fourier transforms of both configurations are given by

\begin{equation}
\phi_s({\bf k}) = L \sum_{n=0}^\infty m_n \left[\delta_{{\bf k},n{\bf
k}_0}+\delta_{{\bf k},-n{\bf k}_0} \right]\label{phik-stripes}
\end{equation}

\begin{equation}
\phi_b({\bf k}) = L \sum_{n=0}^\infty m_{n} \sum_{i=1}^3 \left[\delta_{{\bf k},n{\bf
q}_i}+\delta_{{\bf k},-n{\bf q}_i} \right]\label{phik-bubbles}
\end{equation}

\noindent Replacing Eqs.(\ref{phik-stripes}) and (\ref{phik-bubbles}) into Eq.(\ref{Hk}), we get the mean field free energy in terms of the infinite set of amplitudes $\{ m_n\}$ and $k_0$, which can be considered as variational parameters. Then, truncating Eqs.(\ref{phik-stripes}) and (\ref{phik-bubbles}) to some maximum order $n_{max}$ (i.e., imposing $m_n=0$ $\forall n> n_{max}$) we obtain variational expressions at different levels of approximation for the stripes and bubbles free energies respectively. The complete expressions and some examples are presented in the supplementary material. Assuming that those are the only possible equilibrium states, we can calculate the equilibrium phase diagram by minimizing and comparing the free energies for each type of solution to the same level of approximation $n_{max}$. Except for a few particular analytical approximations, in most of the cases free energy minimization was done numerically using a generalized simulated annealing optimization procedure\cite{TsSt1996}.  All the numerical results shown in this section were done for $b=1$, $u=1$ and $c/k_m=0.1$, which appropriately represents the spectrum of a two dimensional Ising model with exchange and dipolar interactions\cite{CaStTa2004}. Some checks performed with different values of the parameters did not show qualitative differences.

\subsection{Zero field solutions: stripe width variation}

In order to check the limits of validity of the present variational method we first analyze the limit of zero magnetic field. In this case the free energy of the bubbles solution is always  larger than the striped one. Also, due to the up-down symmetry of the system, all the harmonics in  Eq.(\ref{phik-stripes}) with $n$ even do not contribute: $m_{n\,even}=0$.

Direct minimization of Eq.(\ref{Hk}) respect to the fields $\phi({\bf k})$ leads to

\begin{equation}
 A(k)  \phi({\bf k}) =- \frac{u}{L^2}
 \sum_{{\bf k}_1} \sum_{{\bf k}_2}
 \; \phi({\bf k}_1)  \phi({\bf
k}_2)  \phi({\bf k}-{\bf k}_1-{\bf k}_2) \label{eqmf}
\end{equation}.

\begin{figure}
\begin{center}
\includegraphics[scale=0.32,angle=-90]{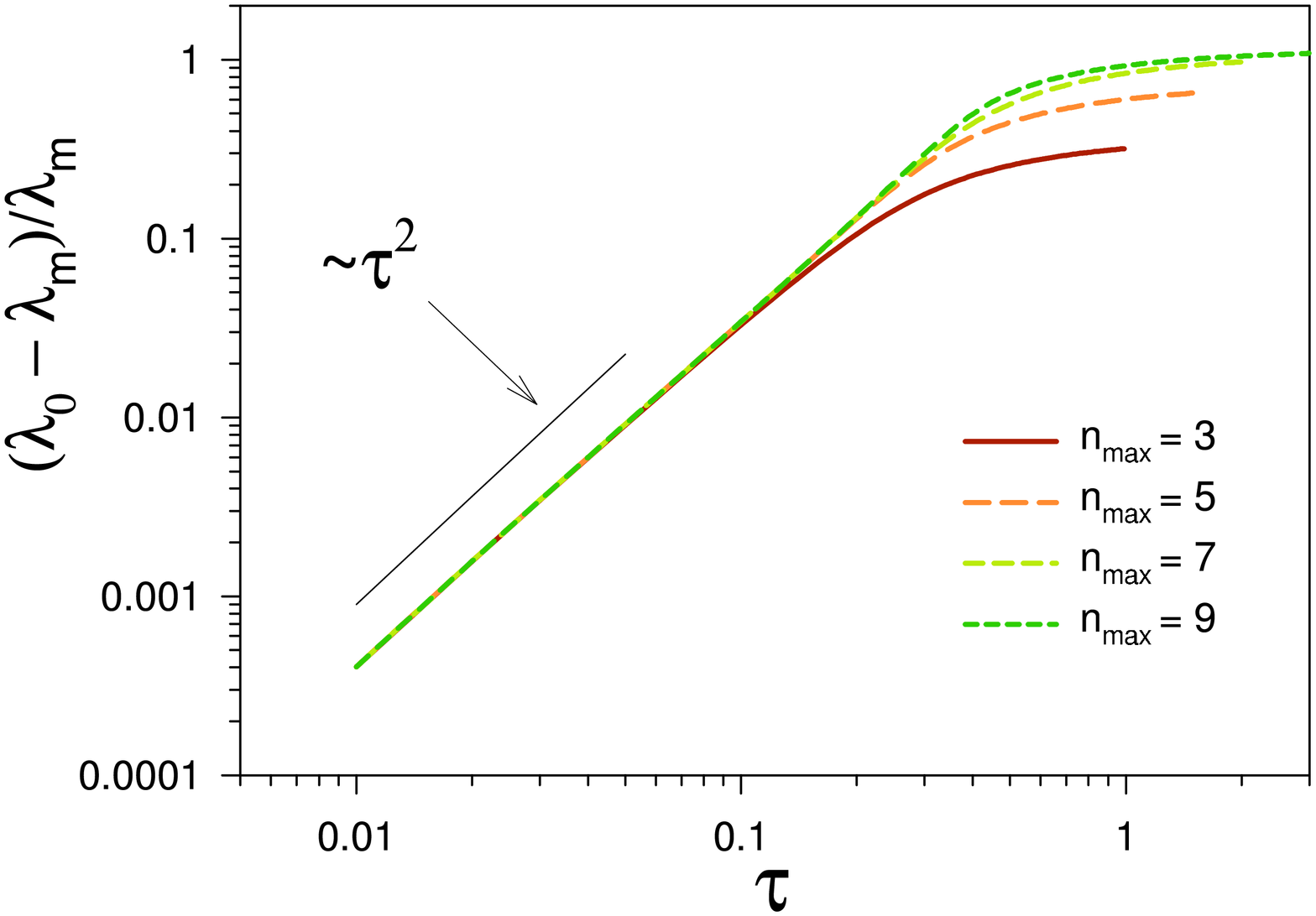}
\caption{(Color online) Stripe width variation with temperature at zero field for  different values of $n_{max}$.\label{lambda}}
\end{center}
\end{figure}

\noindent Replacing Eqs.(\ref{eqmf}) into Eq.(\ref{Hk}) we obtain

\begin{equation}\label{Hmean-field}
F = \frac{1}{4} \sum_{\bf k} A(k) \left| \phi_0({\bf
k})\right|^2
\end{equation}

\noindent where $\phi_0({\bf k})$  are the solutions of Eqs.(\ref{eqmf}). When $T>T_c$ we have $A(k)>0$ $\forall k$ and therefore, from Eq.(\ref{Hmean-field}), the minimum free energy corresponds to the paramagnetic solution $\phi_0({\bf k})=0$ $\forall k$. When $T=T_c$, $A(k_m)=0$ and a single mode solution
with wave vector $k_0=k_m=\pi/\lambda_m$ becomes the minimum of the free energy. As the temperature is further decreased new modes will contribute to the
minimum free energy solution and the stripe width $\lambda_0~\equiv~ \pi/k_0$ is expected to increase.  We performed a numerical optimization considering
up to five modes, i.e., up to $n_{max}=9$. The resulting stripe width $\lambda_0-\lambda_m$ as a function of temperature is shown in Fig.\ref{lambda}.
The results become almost independent of $n_{max}$ for $n_{max}>9$.

For temperatures close enough of to $T_c$ we see that $\lambda_0- \lambda_m \sim \tau^2$, in qualitative agreement with the Bragg-Williams approximation of the dipolar frustrated Ising model\cite{PoVaPe2006,ViSaPoPePo2008}  and also with experimental results on Fe on Cu(100) ultrathin magnetic films\cite{PoVaPe2006}. However, at variance with the dipolar frustrated Ising model, where a crossover to a faster increasing regime (probably exponential) is expected as the temperature decreases, in the present model the stripe width crosses over to a saturation regime. Hence, we can take the crossover value of the reduced temperature $\tau \approx 0.4$ as the limit of validity of the present  approximation, in the sense of reproducing the behavior close to $T_c$ of a more accurate microscopic model.

\begin{figure}
\begin{center}
\includegraphics[scale=0.35,angle=-90]{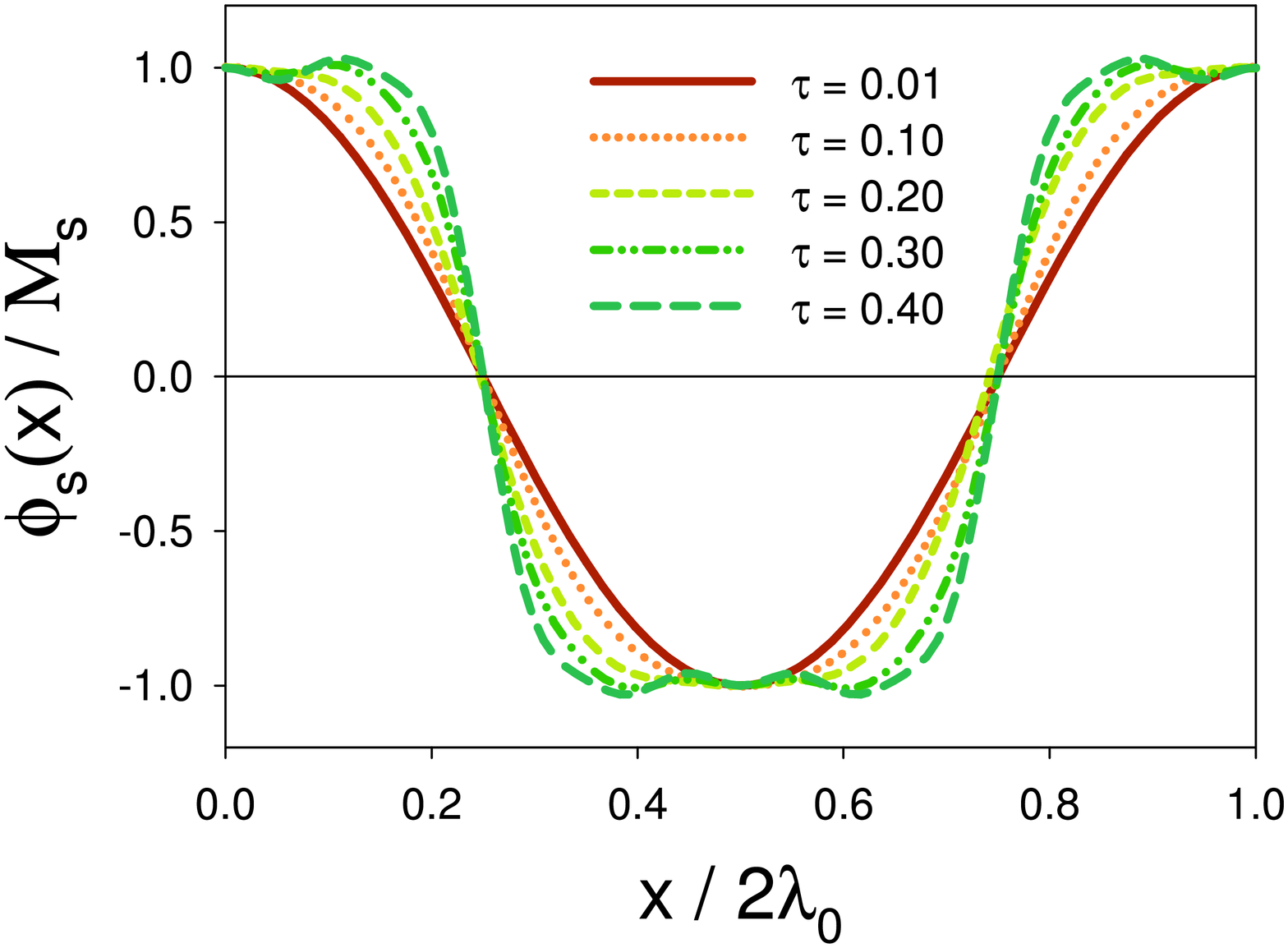}
\caption{(Color online) Change in the magnetization profile with temperature at zero field for $n_{max}=9$. \label{profile}}
\end{center}
\end{figure}

In Fig.\ref{profile} we show the change in the normalized magnetization profile $\phi_s({\bf x})/M_s(T)$, where the saturation magnetization inside a domain is defined as

\begin{equation}
    M_s(T) = \phi({\bf x}=0) = 2\, \sum_{n=1}^{n_{max}} m_n.
\end{equation}

\noindent We see how the profile changes from a sinusoidal shape close to $T_c$ towards a sharp wall type (i.e., square wave form) as the temperature decreases. The same qualitative behavior is observed both in the Bragg-Williams solution of the dipolar frustrated Ising model and experimentally\cite{ViSaPoPePo2008}.

\subsection{Non zero field solutions: phase diagram}

\begin{figure}[h]
\begin{center}
\includegraphics[scale=0.35,angle=-90]{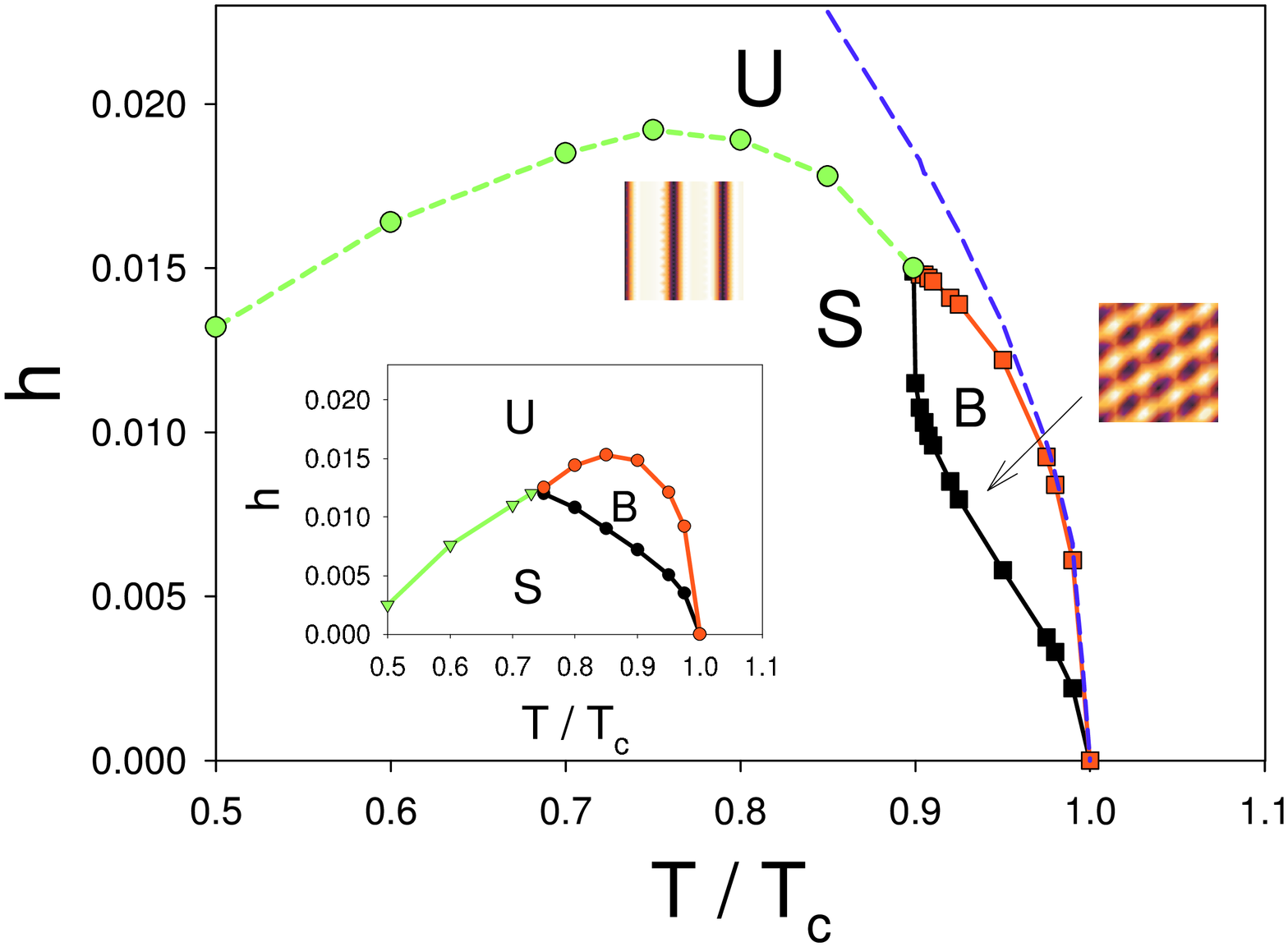}
\caption{(Color online) Phase diagram in the three modes $n_{max}=3$ approximation. Some typical configurations at the stripes and bubble  phases are shown. The blue dashed line correspond to the S spinodal line. The upper B spinodal line lies slightly below the S spinodal line, but both lines are indistinguishable at the present scales. The inset shows the phase diagram in the one mode $n_{max}~=~1$ approximation.\label{diagram2}}
\end{center}
\end{figure}

The minimum level of approximation in this case is to consider only one mode, i.e., only $m_0$ and $m_1$ in Eqs.(\ref{phik-stripes}) and (\ref{phik-bubbles}) are taken different from zero. It is straightforward to verify that minimization conditions in this case imply $k_0=k_m$ for any value of the temperature. In other words, at this level of approximation no stripe width variation with temperature is observed. The phase diagram under this approximation was first obtained by Garel and Doniach\cite{GaDo1982} and later by Andelman {\it et al.}\cite{AnBrJo1987} (see supplementary material). We reproduce such phase diagram in the inset of  Fig.\ref{diagram2} for later comparison with an improved phase diagram. All the transition lines are first order. The most remarkable fact is the absence of a bubbles phase (B) for low enough temperatures, namely at low temperatures there is a single transition line from the stripes (S) to the uniform (U) phase that joins in a triple point with the   S-B and B-U transition lines located close to $T_c$. Our single mode phase diagram differs from the originally
obtained in Ref.~\onlinecite{GaDo1982} (and widely accepted in the literature) but agrees with that from Ref.~\onlinecite{AnBrJo1987}. Interestingly, as
mentioned in the Introduction, already at a single mode approximation an inverse transition is obtained, in qualitative
agreement with recent experiments. Also note that the triple point, at which the bubbles phase cease to minimize the
mean field free energy, corresponds to $T/T_c \sim 0.75$ ($\tau \sim 0.25$), well inside the limit of validity of the continuum model as
discussed in the previous section. Of course, the single mode approximation is probably not reliable at this temperature.

To improve such phase diagram we considered a three modes approximation $n_{max}=3$, namely, we numerically minimized the stripes and bubbles solutions (\ref{phik-stripes}) and (\ref{phik-bubbles}) with respect to the set of parameters $\{k_0,m_0,m_1,m_2,m_3\}$.
Those minimal free energies were compared between them and against the uniform solution (U) from Eq.(\ref{uniform}).
 As in the zero field case, the width of the stripes solution exhibit a quadratic variation with temperature for any value of $h$.
We also verified that the inclusion of further modes in the solutions decreases the free energy for all values of $T$ and $h$.
The resulting phase diagram is shown in Fig.\ref{diagram2}.
The main differences with the one mode phase diagram are the shift of the maximum from the B-U to the S-U transition lines and the shrinking of the
B region with the corresponding movement of the triple point towards the critical temperature. Notice that the location of both the maximum in the
S-U line ($\tau\sim 0.25$) and the triple point ($\tau\sim 0.1$) falls inside the estimated temperature range of validity of the model.  We also observed a strong metastability of both
the S and B phases in all the analyzed temperature range at high values of $h$. The  spinodal line for the S phase is  shown in Fig.\ref{diagram2}. The upper spinodal line for the B phase lies slightly below the previous one (both lines are indistinguishable on the scales of the figure).  Both  the U and B phases are metastable for $T<T_c$ at any value of $h>0$ below the transition lines.

\begin{figure}
\begin{center}
\includegraphics[scale=0.35,angle=-90]{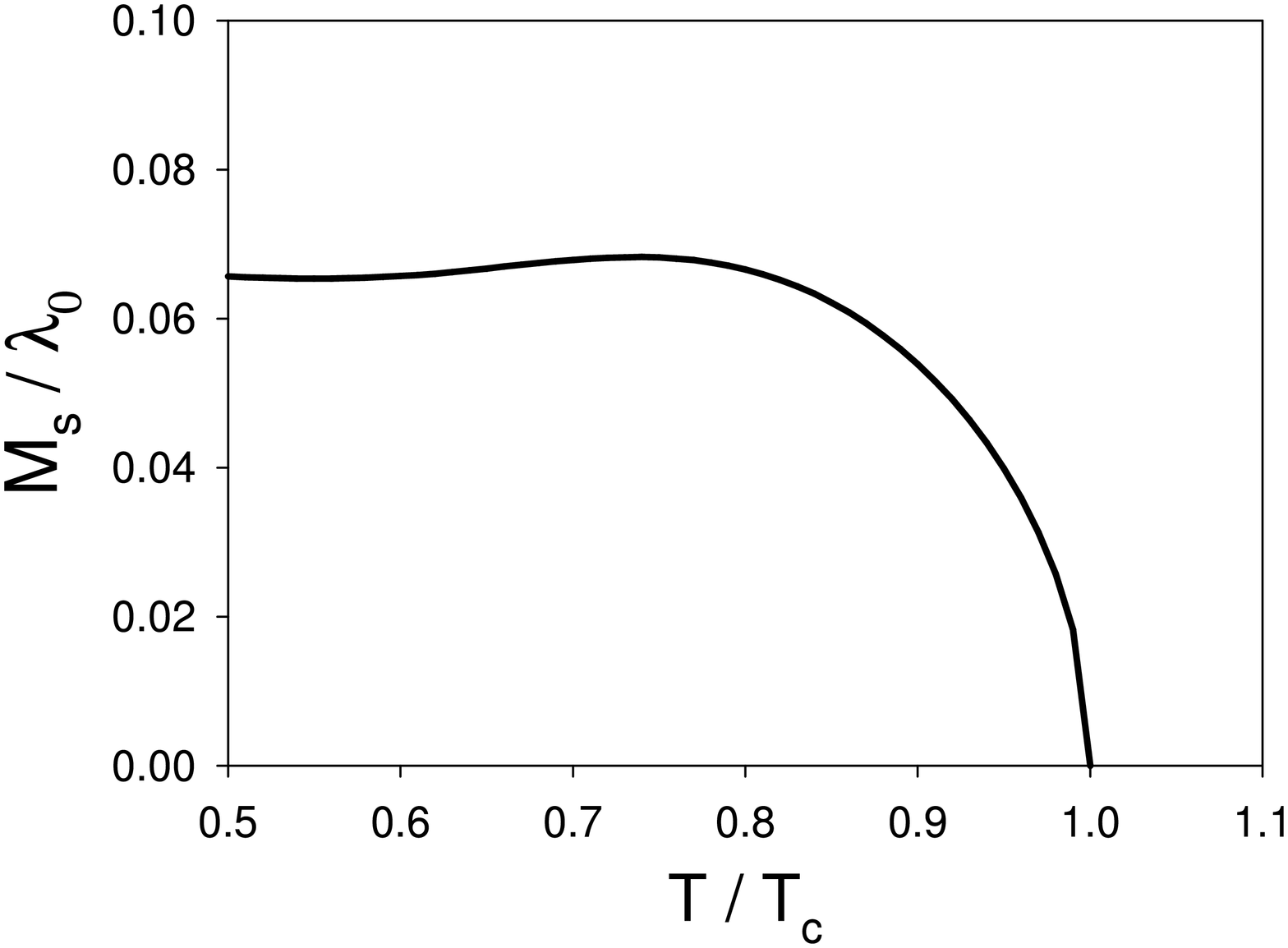}
\caption{\label{hypotesis} Ratio $M_s/\lambda_0$ at zero field as a function of temperature for $n_{max}=9$.}
\end{center}
\end{figure}

Finally, to check the scaling hypothesis proposed in Ref.~\onlinecite{PoGoSaBiPeVi2010} we calculated the ratio $M_s/\lambda_0$ at zero field as a function of temperature, from the results of the previous section. Consistently with the scaling hypothesis,  we see from Fig.\ref{hypotesis} that this ratio exhibit a maximum around $T/T_c\sim 0.75$, like the S-U transition line in the phase diagram from Fig.\ref{diagram2}. As we lower the temperature the ratio increases again because the modulation length saturates while $M_s$ monotonously increases, in an unphysical way.

\section{Monte Carlo simulations in a dipolar Heisenberg model}
\label{mc}

In order to compare the mean field results with the behavior of a specific microscopic model we performed Monte Carlo (MC) simulations using a Heisenberg model with exchange and dipolar interactions, as well as uniaxial out of plane anisotropy. The model, which describes an ultrathin magnetic film (see Ref.\onlinecite{CaBiPiCaStTa2008} and references therein) can be characterized by the dimensionless Hamiltonian:

\begin{widetext}
\begin{equation}
{\cal H} = -\delta \sum_{<i,j>} \vec{S}_i \cdot \vec{S}_j +
\sum_{(i,j)} \left[ \frac{\vec{S}_i \cdot \vec{S}_j }{r_{ij}^3} - 3 \,
\frac{(\vec{S}_i \cdot \vec{r}_{ij}) \; (\vec{S}_j \cdot \vec{r}_{ij})}{r_{ij}^5} \right]
- \eta \sum_{i} (S_i^z)^2
\label{hamiltoniano}
\end{equation}
\end{widetext}

\noindent where the exchange and anisotropy constants are normalized relative to the dipolar coupling
constant, $<i,j>$ stands for a sum over nearest neighbors pairs of sites in a square lattice with $N = L \times L$ sites, $(i,j)$ stands for a sum over {\it all distinct} pairs and $r_{ij}\equiv |\vec{r}_i - \vec{r}_j|$ is the distance between spins $i$ and $j$.
All the simulations were done using the Metropolis algorithm and periodic boundary conditions
were imposed on the lattice by means of the Ewald sums technique. For intermediate values of the anisotropy $\eta$ this model exhibits an in plane - out of plane  reorientation transition\cite{CaBiPiCaStTa2008} at zero magnetic field. In the large $\eta$ limit and for low temperatures, the local magnetization inside the domains is mainly in the out of plane direction,
%with sharp domain walls\cite{PiBiCaSt2009}
and therefore the behavior is expected to be comparable to the scalar Landau-Ginzburg model of the previous section. We choose $\delta=3$ and $\eta=8$. For this set of parameters the system is far away from the reorientation transition, and there is a direct phase transition from the striped state to a perpendicular disordered state at $T_c=1.13$\cite{CaBiPiCaStTa2008}.
 The simulations were carried out for $L=40$. To characterize the magnetic states we calculate the
out-of-plane magnetization:
\begin{equation}
 M_z \equiv \frac{1}{N}\sum_{\vec{r}} \left< S^z(\vec{r}) \right>,
\label{mz}
\end{equation}

\noindent and the orientational order parameter~\cite{CaBiPiCaStTa2008}:
\begin{equation}
 O_{hv} \equiv \left< \left| \frac{n_h-n_v}{n_h+n_v} \right| \right>
\end{equation}
\noindent where $\left< \cdots \right>$ stands for a thermal average, $n_h$ ($n_v$) is the number of horizontal (vertical)  pairs of nearest neighbor spins with antialigned perpendicular component, {\it i.e.},
\begin{equation}
 n_h = \frac{1}{2}\sum_{\vec{r}} \,\left\{1-sig\left[S^z(r_x,r_y),\,S^z(r_x+1,r_y)\right]\right\}
\end{equation}
\noindent and a similar definition for $n_v$, where $sig(x,y)$ is the sign of the product of $x$ and
$y$.

\begin{figure}
\begin{center}
\includegraphics[scale=0.27]{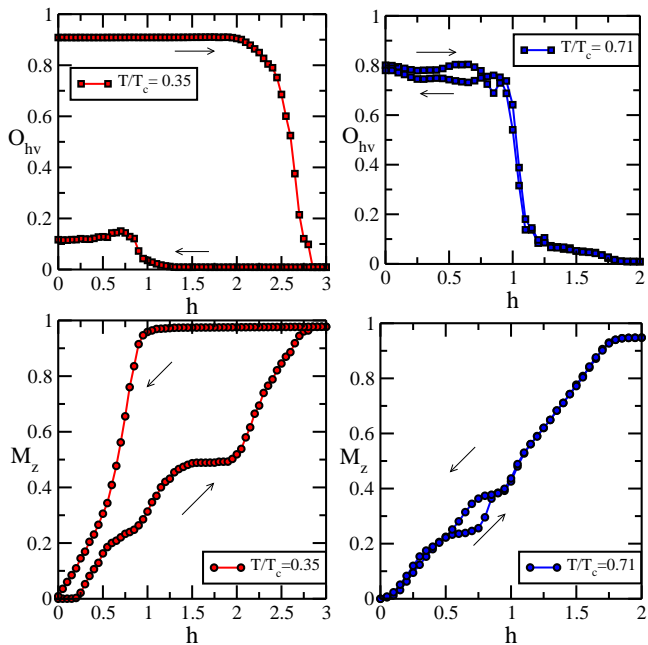}
\caption{\label{cycles} (Color online) Cycles of orientational order parameter $O_{hv}$ (upper panels) and perpendicular magnetization
$M_z$ (lower panels) as a function of field $h$, for two temperatures: $T/T_c=0.35$ (left panels) and $T/T_c=0.71$ (right panels). The arrows indicate the sense of field variation in the cycle.}
\end{center}
\end{figure}

We performed field cycles at fixed temperatures according to the following protocol.
The system is initialized at a stripes configuration (the ground state at $h=0$). The field is increased from zero to a $h_{max}$ and then decreased back to zero, using a ladder procedure with a step variation $\Delta h=0.05$.
At each fixed field the system is thermalized during $t_e=5\times 10^5$ Monte Carlo Step (MCS) and then we calculate
 averages over the next $t_m=10^4$ MCS.  We also performed similar temperature cycles  at fixed field.
Using these cycles together with a visual inspection of the corresponding spin configuration we obtained the $h-T$ phase diagram.

In Fig. \ref{cycles} we show the typical (field cycling) behavior of $O_{hv}$ and $M_z$ for temperatures close to and far away from $T_c$,
namely  $T/T_c=0.35$ and $T/T_c=0.71$. The cycles for $T/T_c=0.35$ show a strong hysteretic behavior, while the curves for $T/T_c=0.71$ show
only a weak hysteresis on a small range of $h$.

In the low temperature regime ($T/T_c=0.35$) we see that $O_{hv}$ drops to zero
at the same field value for which $M_z$ saturates into $M_z=1$, signaling the transition into an uniform state.
As the field increases we see the presence of small plateaus in the magnetization followed by sudden jumps. These jumps correspond to an increase
in the width of stripes with spins aligned to the field preserving the orientational order, as evidenced in the behavior of $O_{hv}$.
Such increase in the size of domains aligned with the external field is in agreement with experimental observations\cite{SaRaViPe2010}.
In this temperature regime no bubbles states are observed and the system goes from a stripes order to a uniform state, with a strong
metastability in both the uniform and the stripes states.

When $T/T_c=0.71$ we observe a clear transition from a striped state to a state without orientational order  at a well defined value of the field $h\sim 1$, without saturation in the magnetization. A visual inspection of the configurations shows indeed
 that, between the stripes order at low field and the uniform state at
hight field, the system displays bubble domains in the range $1 \lesssim h \lesssim 1.75$. Using this criterion we calculate the field values for the stripes-bubbles and bubbles uniform transitions for each temperature.

\begin{figure}
\begin{center}
\includegraphics[scale=0.35]{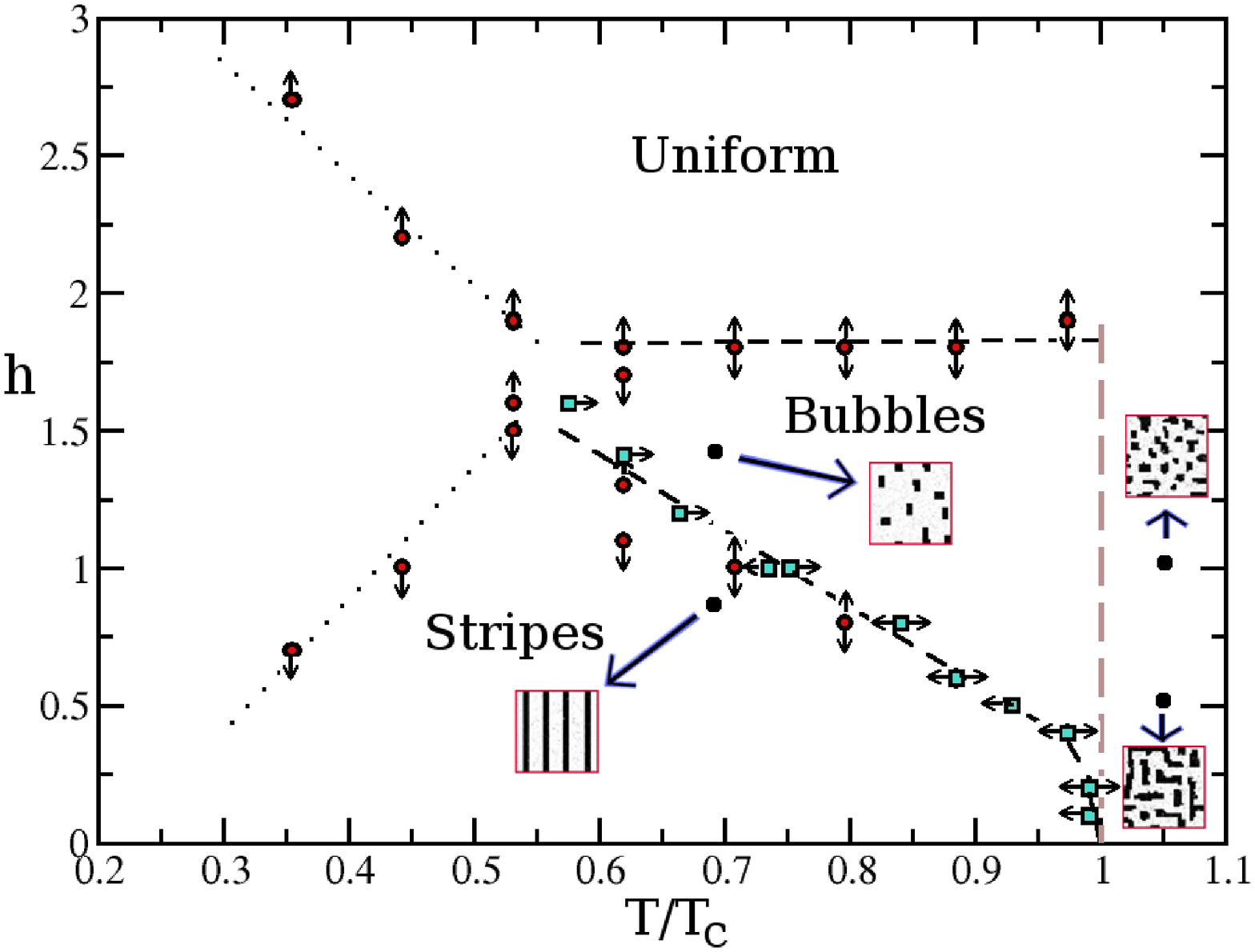}
\caption{\label{diag_eta8} (Color online) MC phase diagram $h$ vs temperature for the dipolar Heisenberg model with $\delta=3$ and $\eta=8$. Different
symbols correspond to different calculation methods of order parameter $O_{hv}$ and perpendicular
magnetization $M_z$: circle for variation in field $h$ at constant temperature and squares for
variation in temperature at constant field. The arrows indicate the directions of parameter variations.
We show some typical spin configuration at different phases. The vertical dashed line is schematic and shows the cross over between regions where bubbles and disordered states can be well differentiated. The dotted lines correspond to the loss of stability (probably spinodal lines) of the stripes (upper line) and uniform (lower line) states when the field is increased (decreased).}
\end{center}
\end{figure}

In Fig.\ref{diag_eta8} we show the phase diagram calculated with the above described procedure. The absence of hysteresis in the temperature range $0.5< T/T_c< 1$ (except very close to $T_c$) allows us to determine with a rather good precision the stripes-bubbles and the bubbles-uniform phase boundaries. Very close to $T_c$ it is very difficult to distinguish between the bubbles state and a uniform state with fluctuations, due to finite size effects. For temperatures above $T_c$ (but close to it) and zero field, the disordered state presents a tetragonal liquid structure\cite{CaMiStTa2006,CaBiPiCaStTa2008}. As the field increases the size of the domains antialigned with the field shrinks, until they break into a disordered arrange (liquid like) of antialigned domains in a ferromagnetic background aligned with the field. Close enough to $T_c$ such state is indistinguishable from a regular arrange of bubbles for small system sizes, both by visual inspection or quantitative  (e.g., structure factor) calculations. A detailed study with larger systems sizes is under way and will be published elsewhere. A similar effect has been observed by MC simulations in the dipolar frustrated Ising model\cite{DiMu2010}. The dashed line in Fig.\ref{diag_eta8} is schematic and shows the crossover region.

On the other hand, at low temperatures ($T/T_c < 0.5$) we did not find any evidence of bubbles states. Although we cannot exclude the possibility of bubbles states hidden by the strong metastability of both the stripes an uniform states, our results appear to be consistent with the existence of a single stripes-uniform first order phase transition at low temperatures, as predicted by the mean field calculation of the previous section.

\section{Conclusions}
\label{conc}

In view of recent theoretical and, specially, experimental results,
the mean field phase diagram of a two dimensional system with competing interactions in an external field
was reconsidered.
Although known from nearly 30 years ago~\cite{GaDo1982}, we have found important differences from the
behavior usually accepted. Interestingly, our results are in qualitative agreement with experimental
results and point once more to the relevance of the mean field behavior in this kind of systems.

First of all, we have found that, already in the single mode approximation, the external field-
temperature phase diagram shows an
inverse symmetry breaking transition between paramagnetic and modulated phases, as recently predicted
based on scaling arguments and confirmed by experiments on ultrathin ferromagnetic films~\cite{PoGoSaBiPeVi2010,SaLiPo2010,SaRaViPe2010}. The presence, and experimental relevance, of the
inverse transitions, was not considered previously in the context of the mean field phase diagram.
From a theoretical point of view, although the generic presence of inverse transitions in this kind
of systems was proposed as a consequence of scaling in the behavior of the characteristic lengths
with temperature~\cite{PoGoSaBiPeVi2010}, we have found that the $h-T$ phase diagram shows a maximum
even when scaling is not expected to occur. This seems to be a very basic property of systems with
competing interactions in external fields.  However, it is worth to stress that, when more accurate solutions (that correctly describe the modulation length variation with temperature) are considered, the scaling hypothesis correctly predicts the location of the maximum in the symmetry breaking transition line between paramagnetic and modulated phases.

Second, we have confirmed that the modulation length strongly depends on temperature below the
paramagnetic-modulated transition, even in the presence of external fields. The quadratic dependence
of the modulation length with temperature is obtained with and without external field. This dependence
is already obtained considering a few harmonics which change the profile of the order parameter from
a simple cosenoidal shape near $T_c$ to a square-like shape as the temperature decreases. At low temperatures
domain walls become sharp and our continuum approximation breaks down.

Third, going beyond the single mode approximation, we have shown that the overall phase diagram does
not change the main characteristics, but the stability lines of the different phases do change. In
particular, the region where the bubbles phase is the thermodynamically stable one shrinks to a small
region near $T_c$ when three modes are considered in the variational solution.

Doing Monte Carlo simulations of an Heisenberg model with perpendicular anisotropy we have found
overall qualitative agreement between simulations and mean field results. In particular, field cycles
at low temperatures show strong hysteretic behavior between stripes and paramagnetic phases, showing
no sign of bubbles.  Although such strong metastability makes it very difficult to determine wether the transition line presents a maximum or not, its existence cannot be excluded from the present results. At higher temperatures hysteresis is greatly suppressed and the saturation is
reached from the low-h stripes phase passing through a bubbles phase at intermediate fields. While
it is difficult to guarantee equilibrium in the system studied, the behavior observed is in general
agreement with mean field predictions of the thermodynamic phases. Large scale simulations would
be very valuable to confirm our preliminary results and the overall picture from the mean field
calculation.

This work was partially supported by grants from
FONCYT/ANPCYT (Argentina), CONICET (Argentina), SeCyT, Universidad Nacional de C\'ordoba
(Argentina) and CNPq (Brazil).

\bibliographystyle{apsrev}
%\bibliography{ultrathin}

\end{document}